\documentclass[useAMS,usenatbib,usegraphicx]{mn2e}
\def\aapr{\ref@jnl{A\&A~Rev.}}		
\graphicspath{ {Figures/} }

\usepackage{amsmath}

\title[The relevance of the Virial ratio]{The difficult early stages of embedded star
  clusters and the importance of the pre-gas expulsion virial ratio} 

\author[J.P.~Farias et al.]{J.P.~Farias$^1$\thanks{E-mail:jfarias@asto-udec.cl}, R. Smith$^{\rm{1,2,3}}$, M. Fellhauer$^1$, S. Goodwin$^4$, G.N. Candlish$^1$,
\newauthor  M. Bla\~na$^{1,5}$, R. Dominguez$^1$ \\ 
  $^1$Departamento de Astronomia, Universidad de Concepcion, Casilla
  160-C, Concepcion, Chile\\ 
  $^2$Yonsei University, Graduate School of Earth System Sciences-Astronomy-Atmospheric Sciences, Yonsei-ro 50, Seoul 120-749,\\ ~Republic of Korea \\
  $^3$Laboratoire AIM Paris-Saclay, CEA/IRFU/SAp, CNRS/INSU, Universit\'e Paris Diderot, 91191 Gif-sur-Yvette Cedex, France\\
  $^4$Department of Physics and Astronomy, University of Sheffield,  Hicks Building, Hounsfield Road, Sheffield, S3 7RH, UK\\
  $^5$Max-Planck-Institut f\"ur extraterrestrische Physik, Gie\ss enbachstra\ss e 1, D-85748 Garching, Germany} 

\begin{document}

\date{Accepted -----. Received -----; in original form -----}

\pagerange{\pageref{firstpage}--\pageref{lastpage}} \pubyear{2013}

\maketitle

\label{firstpage}

\begin{abstract}
We examine the effects of gas-expulsion on initially
substructured distributions of stars. We perform N-body simulations 
of the evolution of these distributions in a static background potential to
mimic the gas. We remove the static potential
instantaneously to model gas-expulsion.  We find that the exact 
dynamical state of the cluster plays a very strong role in affecting a cluster's
survival, especially at early times: they may be entirely destroyed or
only weakly affected. We show that knowing both detailed dynamics and
relative star-gas distributions can provide a good estimate of the
post-gas expulsion state of the cluster, but even knowing these is not
an absolute way of determining the survival or otherwise of the cluster.
\end{abstract}


\section{Introduction}
\label{sec:intro}

The vast majority of stars appears to form in environments with
densities typically much greater than the field 
\citep{Lada2003,Bressert2010,King2012}, but after a few Myr the majority
of the stars are dispersed into the field \citep{Lada2003}.  The
mechanism by which stars are dispersed is unclear, they may form
unbound, or form in bound clusters which are then unbound by the
expulsion of residual gas left-over after star formation (see below).

Star formation does not consume all of the gas in a molecular cloud,
indeed it is estimated that at most 30 per cent of the gas is turned
into stars \citep{Dobbs2014,Padoan2014}.  Observations show that by 10~Myr, and probably
well before, young stars are no longer associated with gas
\citep{Lada2003}.  This gas has presumably been heated and expelled by
feedback (ionisation, mechanical winds and supernovae).  It is 
interesting and important to examine how the stars respond to this gas
loss and the corresponding (very significant?) change in the local
potential.  

Many authors have examined the effects of gas loss on stars (see
\cite{Tutukov1978,Hills1980,Mathieu1983,Elmegreen1983,Lada1984,Elmegreen1985,
Pinto1987,Verschueren1989,Goodwin1997a,Goodwin1997b,Geyer2001,Boily2003a,Boily2003b,Goodwin2006,Bastian2006,Baumgardt2007,Parmentier2008,Goodwin2009}),
but the majority of this work has concentrated on gas loss from
clusters in which the stars and gas are both dynamically relaxed and
in global virial equilibrium (but see
\cite{Lada1984,Verschueren1989,Goodwin2009}).  If one assumes that the
gas and stars in the cluster are well mixed and relaxed then the
gas-to-star mass ratio is enough to derive the  global star formation efficiency (SFE) in the cluster.
However, \cite{Verschueren1989}
and \cite{Goodwin2009} note that the exact
dynamical state of clusters at the moment of gas expulsion is also
extremely important and point-out that
the SFE alone is not the most important factor in deciding the fate of
a cluster.

Recently it has become very clear that not all stars form in relaxed,
centrally concentrated clusters, and can often form in complex
hierarchical/substructured distributions which follow the gas 
\citep{Whitmore1999,Johnstone2000,Kirk2007,Gutermuth2009,Difrancesco2010,Maury2011,Konyves2010,Wright2014,Schmeja2008}.
It is still unclear which of clustered or
`hierarchical' is the main mode of star formation (or how and if they
are connected), but it is clear that
hierarchical is an important mode in many nearby low-mass star forming
regions.

In a hierarchical mode of star formation the stars and gas are not in
anything close to equilibrium. Both will be in motion relative to one another
as the stars respond only to gravity, where as the gas can suffer additional 
hydrodynamical effects and feedback.  This
means that the absolute and relative distributions of stars and gas at 
formation can change significantly before the gas is removed from the system.  
Therefore the effect of gas removal on the stellar distribution depends not just
on the relative masses of the stellar and gaseous components, but also
on how they have dynamically evolved.

This is the latest in a series of papers in which we have examined the
response of complex, hierarchical systems to gas expulsion (see
\cite{Smith2011a,Smith2013a}).  The complexity of hierarchical distributions
means that there is a very large parameter space to explore, and a
wide variety of possible outcomes.  In this paper we expand on the
work of \citet{Smith2011a,Smith2013a}.

\section{Simulations}
  \label{setup}

Since we wish to continue the work of \citet{Smith2011a,Smith2013a},
we use similar, simplified initial conditions for our 
simulations.  We perform N-body simulations using the Nbody6 code
\citep{Aarseth2003}.  

As we describe in more detail below, equal-mass stars are distributed in a fractal
distribution in a smooth and static background
potential to mimic the potential of the gas in which they are embedded. 
Given that we use a static potential for the gas, we are unable to include active star formation in 
our models. However we do not expect this to change the key conclusions of this study.
The potential is then removed instantaneously to
simulate gas-expulsion. 

This is clearly an extreme over-simplification in many ways.  In reality, the gas
is not distributed in a smooth spherical distribution, and both the
gas and stars will move in response to changes in the global
potential. The gas will also react to hydrodynamic forces and
feedback (which is what eventually expel any remaining gas).  Gas
expulsion is unlikely to be instantaneous, rather gas will be lost at
different rates in different regions, and dynamics can cause the stars
and gas to decouple without any feedback.

We take this simplified approach rather than attempt to deal with the
gas dynamics with a hydrodynamic method for two reasons.  Firstly, the
practical issue of performing large ensembles of simulations -- this
is much quicker and easier with pure $N$-body simulations. Secondly,
the complexity of the gas distribution would add large numbers
of (largely unknown) parameters to our possible parameter space.
We will return to discuss this issue later.

We choose equal mass
particles in order to avoid complex two body interactions and mass
segregation (see e.g. \citealt{Allison2009} for the complications a
realistic mass spectrum can add to an already complex
problem). This will be addressed in more detail in a future paper 
(Bla\~na et al., in prep.).

\subsection{Initial distributions}

In all cases we model the stellar distribution using $N = 1000$
particles with equal masses of 0.5~M$_{\odot}$.  

Using the box fractal method described by \citet{Goodwin2004}, we
create 20 random realisations of fractal distributions, each with a fractal dimensions of $D =1.6$, corresponding to a highly clumpy initial distribution within a
radius of 1.5~pc.  We use the same 20 stellar distributions for each background potential.

We start our simulations with two energies: 
initial virial ratios of $Q_{\rm i} = 0.5$ (warm), and  $Q_{\rm i} =
0$ (cold). As we will show, even our $Q_{\rm i=0.5}$ simulations are not in equilibrium. 
Fractal clusters will then attempt to relax in persuit of equilibrium and subsequently there are 
large variations in their virial ratio parameter. Thus, we measure $Q$ instantaneously at
two important epochs: the beginning of the simulation ($Q_{\rm i}$ where `i' donates `initial'),
and the moment when gas expulsion begins ($Q_{\rm f}$ where `f' donates `final'). The cold systems start with the stars initially at rest relative to each-other, this is unrealistic, but is the case
where we expect the most rapid collapse and erasure of substructure.

\subsection{The background potential and SFE}

We work with three different static background potentials: (i) a Plummer sphere with $R_{\rm pl} = 1.0$ pc and
$M_{\rm gas,tot} = 3472$~M$_{\odot}$, (ii) a uniform sphere of gas with a maximum 
radius of $R = 1.8$~pc and $M_{\rm gas,tot} = 3455 \rm~M_{\odot}$ (equivalent to a Plummer sphere with $R_{\rm pl} =\infty$), and finally (iii) a highly concentrated Plummer sphere with $R_{\rm pl} = 0.2$ and $M_{\rm gas,tot} = 2053$~M$_{\odot}$. This choice of parameters ensures that we obtain a $\rm
SFE = 0.2$ for all three background potentials (i.e. we always have
exactly 2500~M$_{\odot}$ total mass within 1.5~pc, of which 2000~M$_{\odot}$ is
gas, and 500~M$_{\odot}$ is stars).

In this work we expel the gas instantaneously at early times in the evolution
of the cluster i.e. within a few crossing times ($1 t_{\rm cr} \approx
1.4$ Myr), and compare to clusters with a later gas expulsion ($\sim
7.5 t_{\rm cr}$).

\subsection{Gas expulsion time}

We simulate rapid gas expulsion by removing the background
potential instantaneously.  This is the most potentially destructive
gas expulsion (see \citealt{Baumgardt2007}).

As we wish to model the effects of early gas expulsion, we usually remove the 
gas potential instantaneously within two {\em initial} crossing times 
($1 t_{\rm cr} \approx 1.4$ Myr). During this time, the initial distributions relax violently 
  and $t_{\rm cr}$ may no longer be a representative timescale (see section \ref{sec:motivation})

We first summarise the results from our previous studies before
describing and explaining our new results.

\subsection{Previous work}

Numerical models of gas expulsion from initially virialised gas-star
Plummer spheres have shown that a small fraction of stars can remain
bound if the stars make-up more than about 30 per cent of the initial
system (in this case what is often assumed to be a direct measure of
the star formation efficiency), and the majority of the stars will
remain bound if the fraction is greater than 50 per cent (see e.g. \citealt{Goodwin2006,Baumgardt2007}).  The speed of gas expulsion
is important with fast (instantaneous) gas expulsion being
significantly more disruptive than slow (adiabatic) gas expulsion (see
\citealt{Baumgardt2007} and  \citealt{Lada1984}).

Our initial conditions are a highly simplified, but hopefully
realistic at a fundamental level, model of a fractal stellar
distribution relaxing in a global gas potential before the removal of
that gas potential.  This is very different from the initially star-gas
equilibrium distributions assumed in most previous studies.

Because the stellar distribution is highly out-of-equilibrium and also
different from the gas potential, this means that the stellar
distribution will violently relax.
The initial fractal substructures will be erased and the stellar
distribution will become smooth, whilst at the same time relaxing to
fit the underlying static (gas) potential (see also \citealt{Allison2009,Parker2014}).
This means that the stellar distribution will become more concentrated
relative to the static gas potential as potential energy stored in
substructure is distributed more smoothly (see e.g. \citealt{Allison2009}).

 \cite{Smith2011a} identify an important parameter in determining the
remaining bound fraction as the local stellar fraction (LSF).  The LSF
is a measure of the gas mass within the {\em stellar} half-mass
radius; i.e. a measure of the relative importance of the gas to the stars.
The LSF is defined as
  \begin{eqnarray}
    \rm LSF & = & \frac{M_{*}(r < r_{\rm h})} {M_*(r < r_{\rm h}) +
    M_{\rm gas}(r < r_{\rm h})} 
    \label{eq:lsf}
  \end{eqnarray} 
where $r_{\rm h}$ is the radius that contains half of the total
mass in stars.  $M_{*}$ and $M_{\rm gas}$ is the mass of stars and gas,
respectively, measured within $r_{\rm h}$.

The LSF is analogous to the star formation efficiency (SFE) quoted in
many previous papers (although as noted by \cite{Verschueren1989} and
\cite{Goodwin2009} this should really be called the effective SFE as
its relationship to the true SFE is uncertain).  \cite{Smith2011a}
show that the LSF will depend on the initial distribution of stars,
the initial gas-to-star mass, and the initial energy of the stellar
distribution \citep[also see][]{Parmentier2013,Parmentier2014}.

 \cite{Smith2011a} find that there is a reasonably good relationship
between the final bound fraction and the LSF at the point of gas
expulsion for systems which have relaxed for more than two initial crossing times.
However, \cite{Smith2013a} show that if gas expulsion occurs earlier, it is rather more complex than
this suggests.

The longer the stars have to relax, the closer to a virialised, smooth
distribution in equilibrium with the static gas potential they will
become.  \cite{Smith2013a} show two important consequences of this
relaxation processes.  Firstly, the LSF changes with time and so the exact time of gas expulsion is very
important.  Secondly, the violent relaxation of the initially clumpy
stellar distributions is stochastic and initial distributions that are
initially `the same' (i.e. drawn from the same generating functions)
can evolve very differently, and at any particular point in time
(i.e. at gas expulsion) can have quite different dynamics and be at
different `stages' in their relaxation.  If gas expulsion occurs at
early times (typically less than one crossing time, or around 1~Myr)
then the LSF ceases to be a good predictor of the final bound fraction.

 \cite{Smith2013a} attempted to quantify these effects and found that
the virial ratio of the stars at the time of gas expulsion is {\em also}
very important to the final bound fraction (as suggested by Goodwin
2009).  In this study we concentrate on examining the effects of the
stellar virial ratio at the time of gas expulsion.

\subsection{Motivation}
\label{sec:motivation}

In this paper we extend the work of \cite{Smith2011a,Smith2013a}.  We
have two related questions that we wish to consider.

Firstly, to what extent is it possible to predict the final bound
fraction of the system?  Secondly, is it ever practically possible (either
observationally or theoretically) to predict the final bound fraction
of a particular system?

In this paper we concentrate mainly on the effects of the dynamical
state of the stars at the time of gas expulsion (as measured by the
stellar virial ratio).

We concentrate on systems which have not had many crossing times to
relax.  For the systems we simulate here the instant of gas expulsion are
typically 1--3~Myr, or less than 2 initial crossing times.  This means that the
initial substructured distributions have not had time to relax and are
in the process of violent relaxation.  It is worth noting that this
corresponds to the age of the gas-free Orion Nebula Cluster \citep{Jeffries2011}.

We will refer to the virial ratios of the systems, $Q = T/|\Omega|$ where $T$ is the
kinetic energy, and $\Omega$ the potential energy.  $Q=0.5$
corresponds to virial equilibrium, but we note that our systems
(especially initially) are often {\em not} in any true equilibrium, even if $Q=0.5$
(they might be formally virialised, but may not have equilibrium
spacial or velocity distributions). Nevertheless as we shall describe below
$Q$ is a very useful measure.

\subsubsection{The evolving dynamical state of the cluster}

At the start of the simulation we have a very out-of-equilibrium
distribution with a $Q_{\rm i} = 0$ or $0.5$.  The stars will immediately
start to violently relax and erase the substructure present in the
system (see also \citealt{Allison2009,Parker2014}).  With our
initial conditions there will always be an initial collapse of most of
the stars.  Violent relaxation rapidly, but very roughly, attempts to
bring the system to a rough dynamical equilibrium (both virial
equilibrium of the energies, and a smooth density field). 

Therefore the stellar component of the system rapidly changes its density
distribution, size, and the way that energy is distributed.  This
means that any {\em initial} measures of size, energy etc. rapidly
change, meaning that any useful timescale such as crossing time also
change.

We take as a measure of the state of the cluster the value of the
virial ratio, $Q$, at any time as well as the rate at which the virial
ratio is changing, $\dot{Q}$.

\begin{figure}
    \begin{center}
      \includegraphics[width=3.35in]{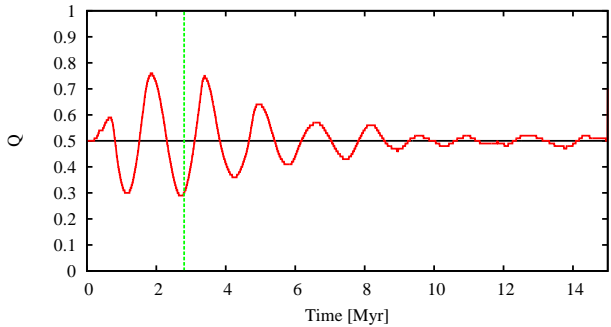}
    \end{center}
    \caption{A representative example of the virial ratio variations
      with time of an out-of-equilibrium distribution of stars
      (a fractal in this case) inside a smooth background potential. As we study early gas expulsion, the smooth background potential is instantaneously removed before two crossing times occur (i.e to the left of the vertical green dashed line).} 
    \label{fig:virialtime}
  \end{figure}

In Fig.~\ref{fig:virialtime} we show the evolution of the virial ratio
with time for a typical system starting with $Q_i = 0.5$.  Even though
this system starts in `virial equilibrium' it immediately increases
its $Q$, and then oscillates around $Q=0.5$ with decreasing amplitude.

What happens is that the cluster immediately starts to violently relax
and attempts to collapse into the gas potential (thus $Q$ rises as
potential energy is converted into kinetic energy in the initial
collapse).  But the initial collapse is soon halted and the stellar
distribution expands causing $Q$ to fall, as the stars oscillate within the potential well of the cluster.  Whilst this is happening
substructure is also being disrupted, and within a few oscillations
the system smooths out and the oscillations represent a `pulsation' of
a smooth cluster as it attempts to fully virialise. Therefore the oscillations in $Q$ with time provide an internal measure of the
level to which the system has relaxed.  

\subsubsection{Gas expulsion times}

When gas expulsion occurs is (yet another) key parameter in setting
the final state of the system as quantified by the final bound
fraction (see \citealt{Goodwin2009,Smith2011a,Smith2013a}).  In our
simulations this is modelled by the time at which we remove the
static background gas potential to represent instantaneous gas
expulsion.  (Obviously this is a huge over-simplification which we
return to in the discussion.)

In \cite{Smith2013a} we showed that the value of the virial ratio at
the start of gas expulsion, $Q_{\rm f}$, is important -- is the system in an
expanding or contracting part of its relaxation process? However in \cite{Smith2013a}, 
we chose a fixed instant in time for gas expulsion for all fractals. 
As each random realisation of a fractal evolves differently, 
the exact virial ratio at the moment of gas expulsion was very varied, and uncontrolled.

In order to better control the dynamical state of the cluster
at the point of gas expulsion, we artificially vary the instant at which gas expulsion occurs 
(between 0--2 crossing times)  so as we can choose the virial ratio of the cluster. The upper 
limit for the time of gas expulsion is marked by the green dashed vertical line in Fig.~\ref{fig:virialtime}. 
For example, in one series of ensembles we ensure that $Q_{\rm f} = 0.5$ by forcing gas expulsion 
to occur whenever the virial ratio happens to be at $Q=0.5$. 

Obviously real systems will not always expel gas at a pre-determined
value of $Q_{\rm f} = 0.5$, so we also expel the gas at other times as
$Q_{\rm f}$ varies from subvirial ($Q_{\rm f} \sim 0.2$) 
to supervirial ($Q_{\rm f} \sim 0.7$).

\subsection{The full set of initial conditions}

To summarise our set of initial conditions:

We take ensembles of 10 or 20 statistically identical systems (all
parameters from the same generating functions) with $N=1000$
equal-mass stars with $M=0.5$M$_\odot$ distributed as a $D=1.6$
fractal with radius 1.5~pc.  The velocities of the stars are scaled to
give initial virial ratios for the stellar system in the background potential of 
$Q_{\rm i}=0$ or $Q_{\rm i}=0.5$.

These stellar distributions sit in a three different static background
potentials. A Plummer sphere with $R_{\rm pl}=1$~pc, a highly concentrated Plummer
sphere with $R_{\rm pl}=0.2$~pc, and a uniform sphere. All of them
ith a total mass of 2500~M$_\odot$ within 1.5~pc that
ensures an effective SFE~=~0.2.

The systems are evolved and their time-evolving virial ratios are tracked.  
The instant of gas expulsion is varied in order to have gas
expulsion ocur when the final stellar virial ratio has a wide range of stellar virial ratios from subvirial
($Q_{\rm f} \sim 0.2$) to supervirial ($Q_{\rm f} \sim 0.7$). At the 
moment of gas expulsion the local star fraction (LSF) can be calculated.

They are then evolved until the simulation reaches 15 Myr ($\sim$10.7 initial
$t_{\rm cr}$) at which the number of
stars still bound in a remaining cluster can be found to give the
final bound fraction, $f_{\rm bound}$.

We reiterate that these are not very `realistic' initial conditions,
and there is much about them that is clearly artificial.  However,
even as artificially simplified as they are, they are still extremely
messy and complicated.  Their use is not to model reality directly,
but rather to allow us to probe the physics behind relaxation and
recovery after gas expulsion.

\begin{figure*}
    \begin{center}
      $\begin{array}{ccc}
        \includegraphics[width=3.2in]{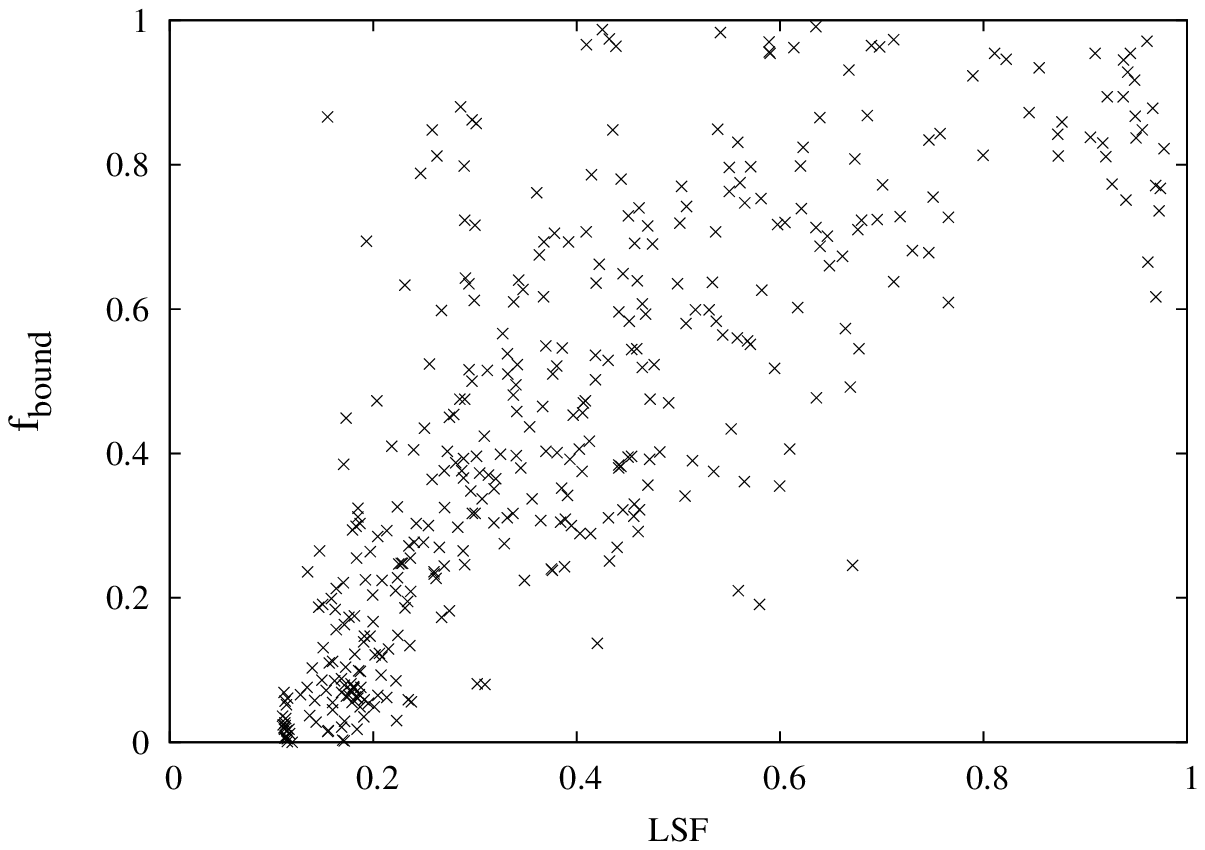} & &
        \includegraphics[width=3.2in]{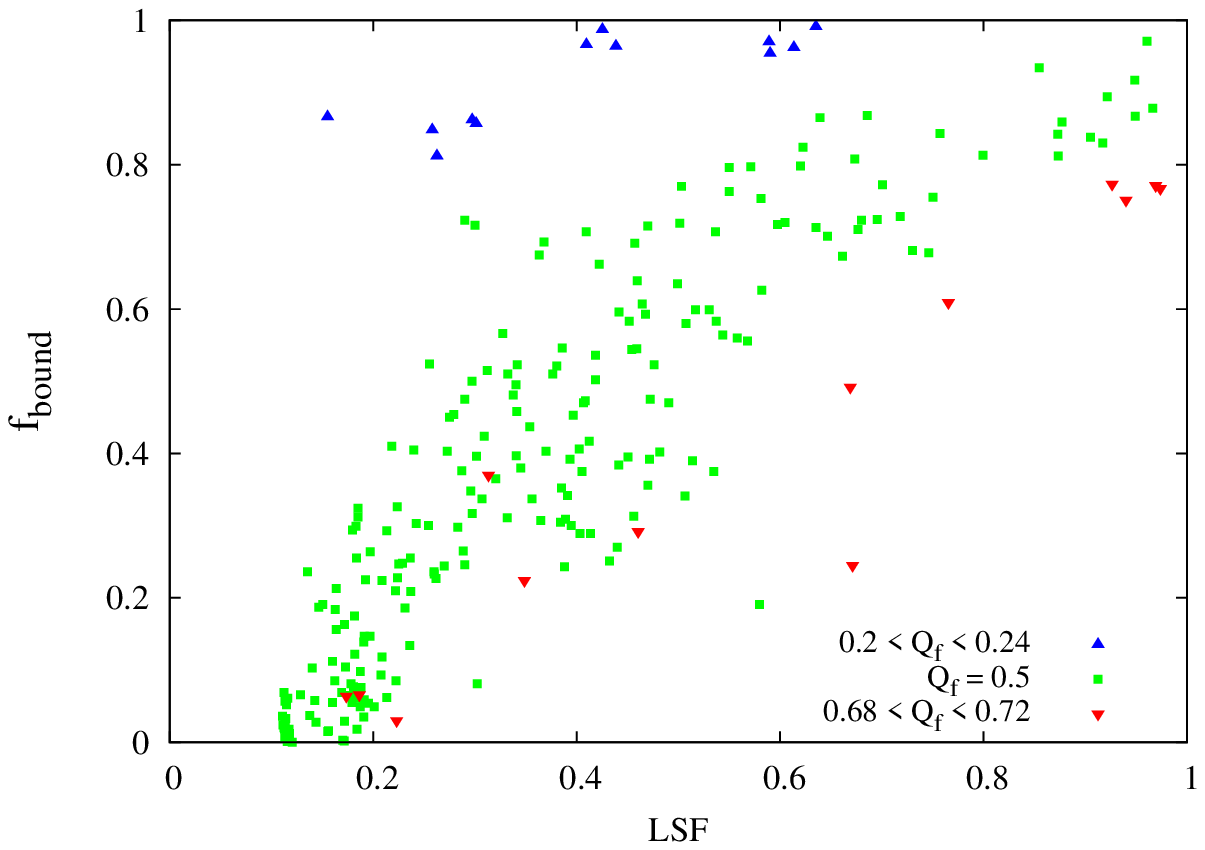} \\
        \text{a} &  & \text{b} \\
      \end{array}$
    \end{center}
    \caption{\textbf{a}: Crosses show $f_{\rm bound}$ against LSF at the time of
      gas expulsion for  
    all the simulations carried out in this study (see text for details).
    \textbf{b}: Simulations that are highly sub-virial ($Q_{\rm f} = 0.22--0.24$; blue triangles), $Q_{\rm f} = $ exactly 0.5 (green squares), or highly super-virial ($Q_{\rm f} = 0.68$--$0.72$; red
  inverted triangles) at the instant of gas expulsion} 
    \label{fig:alldata}
  \end{figure*} 

\section{Results}

The key parameter that we wish to investigate is the fraction of stars
that remain in a bound cluster after gas expulsion and the post-gas
expulsion relaxation of the system.  This `bound fraction' ($f_{\rm
  bound}$) is the size of the naked cluster that remains.
To measure the bound fraction we use the 'snowballing method'. 
In this technique particles that are bound to the cluster are found in 
an iterative procedure, that corrects for the systemic
velocity of the cluster at each iteration. (see sec. 2.6 of \citealt{Smith2013b} for a more complete description).

\subsection{Final bound fractions}

  In Figure~\ref{fig:alldata}a we show the final bound fraction, $f_{\rm bound}$, against
the local star fraction, LSF, for {\em all} the simulations we have
run in this paper.

There is a vague trend that a high-LSF results in a high-$f_{\rm
bound}$ (i.e. the bottom right corner of Figure~\ref{fig:alldata}a is empty).  
But there is a huge amount of scatter in this figure, in particular around LSF of 0.2
can result in clusters with an $f_{\rm bound}$ between zero and almost
unity.  For any particular value of LSF there is a scatter of at least
0.5 in $f_{\rm bound}$.

This might suggest that there is no way of estimating the final bound
fractions of star clusters after gas expulsion.  We show below that
it is possible to understand the system and fairly accurately predict
the final bound fractions if one knows both the LSF and stellar virial
ratios at the time of gas expulsion. 

Because of how we have (rather artificially) chosen our gas
expulsion times we can split the simulations shown in Figure~\ref{fig:alldata}a into
groups depending on their final virial ratios.  In Figure~\ref{fig:alldata}b we only plot
the simulations with $0.22<Q_{\rm f}<0.24$ (blue), $Q_{\rm
  f} \sim 0.5$ (green), and $0.68<Q_{\rm f}<0.72$ (red).

  It is clear from Figure~\ref{fig:alldata}b that a significant amount of the scatter is
due to the value of $Q_{\rm f}$ at the time of gas expulsion.  The
$0.22<Q_{\rm f}<0.24$ simulations all have $f_{\rm bound} \sim 1$.
The $Q_{\rm f} \sim 0.5$ simulations show a rapid increase in $f_{\rm
  bound}$ with LSF for low-LSF, then a very roughly linear increase.
And the $0.68<Q_{\rm f}<0.72$ simulations show a roughly linear increase
in $f_{\rm bound}$ with increasing LSF.

\subsection{A simple physical model}
\label{sec:theo}

\begin{figure*}
    \begin{center}
      \includegraphics[width=6.3 in]{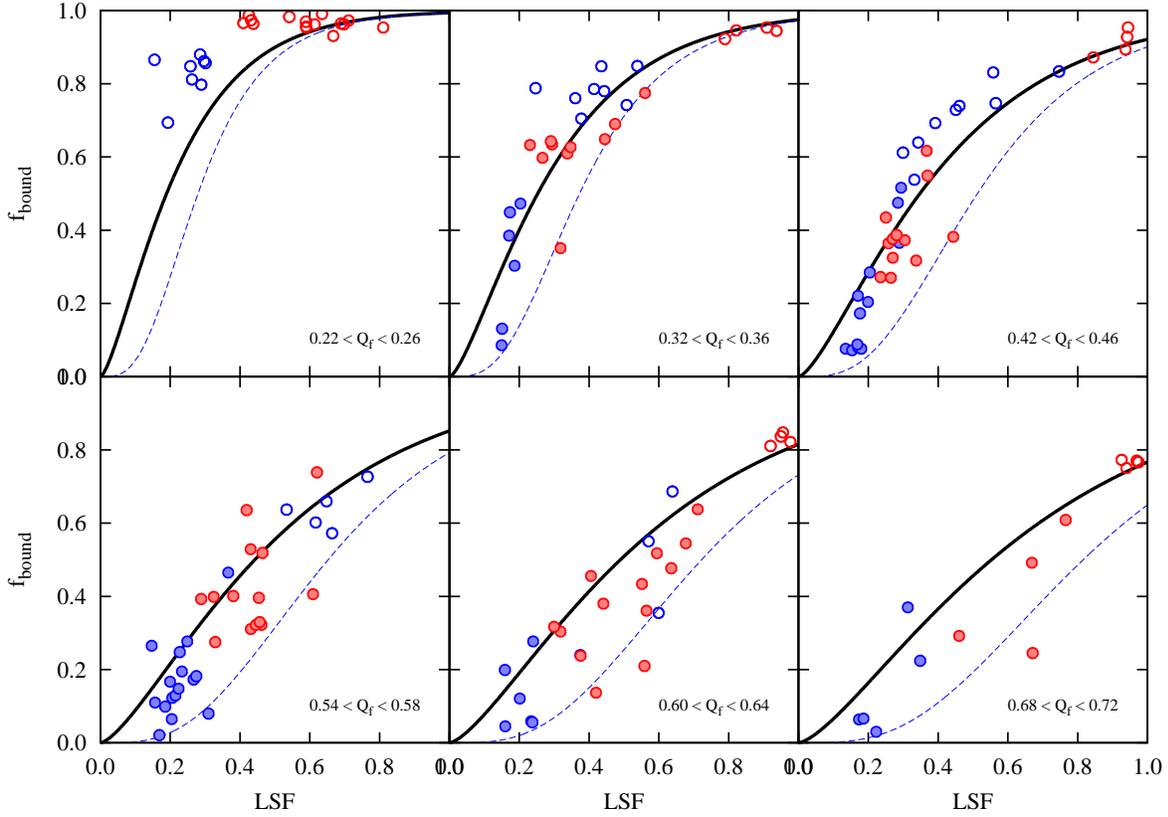}
    \end{center}
    \caption{The $f_{\rm bound}$-LSF trend for different virial ratios. Colors
  represent the shape of the background gas, a Plummer sphere (blue) and a uniform
  sphere (red), filled circles are simulations with $Q_{\rm i}=0.5$ and open
  circles are distributions with $Q_{\rm i}=0.0$. The black solid
  lines and blue dashed lines are the fits from the model
  described in sec.~\ref{sec:theo}. }
    \label{fig:extremes}
  \end{figure*}

In Fig.~\ref{fig:extremes} we plot $f_{\rm bound}$ against LSF for
bins of different $Q_{\rm f}$ increasing from low-$Q_{\rm f}$ in the
top left to high-$Q_{\rm f}$ in the bottom right.  Systems with
initial virial ratios of $Q_{\rm i}=0.5$ are marked by filled circles,
those with $Q_{\rm i}=0$ by open circles.  

The black solid lines and blue dashed lines are a simple model fit to
the data which we describe in this section.   Note that the colours
show the form of the gas potential which we will describe in the next
subsection.  For now we will concentrate on building a simple model to
fit the $f_{\rm bound}$ against LSF trends with different $Q_{\rm f}$.

We can construct a very simple analytical model that fits the results
of our simulations surprisingly well (see \citealt{Boily2003a} for a
similar, but rather more detailed derivation).

As described above and in previous papers, the initial fractal stellar
distribution will attempt to relax and virialise within the gas
potential.  What are important for the impact of gas expulsion are two
quantities {\it at the time of gas expulsion}: the virial ratio $Q_{\rm
f}$ of the stars relative to the gas and the local stellar fraction
LSF.  The LSF measures the relative masses of the gas and the stars
within the stellar half-mass radius (see above).  Therefore the total
mass (stars plus gas) $M_{\rm tot}$ in the region in which the stars
are present is $M_{\rm tot} \sim M_{*}/{\rm LSF}$.

Let us denote quantities just before the gas expulsion with index 1
and just after the gas expulsion with 2.

One quantity of interest is the kinetic energy $T_{*}$ of the stars,
set by their velocity dispersion $\sigma_{*}$. If we assume a Maxwellian velocity 
distribution, the kinetic energy is given by:

\begin{eqnarray}
  \label{eq:ekin0}
  T_{*,1} & = & \frac{3}{2} M_{*} a^{2}
\end{eqnarray}

where $a$ is the scale factor of the Maxwellian velocity distribution. $a$ is
related to the velocity dispersion as $a^2=\sigma_*^2\pi/(3\pi-8)$. Therefore,
\begin{eqnarray}
  \label{eq:ekin1}
  T_{*,1} & = & \frac{3 \kappa }{2} M_{*} \sigma_{*}^{2}
\end{eqnarray}

where $\kappa=\pi/(3\pi+8)$. After gas expulsion the stars have
not had time to change their kinetic energy (since the gas is expelled instantaneously) and 
so we can assume $T_{*,2} = T_{*,1}$.

The potential energy of the stars before gas expulsion can be
approximated by 
\begin{eqnarray}
  \label{eq:epot1}
  \Omega_{*,1} & \sim & - M_{*} \frac{GM_{\rm tot}}{r_{\rm h}}
\end{eqnarray}
were $G$  is Newton's gravitational constant, while the potential energy after the gas is lost is only due to the
potential made by the stars alone
\begin{eqnarray}
  \label{eq:epot2}
  \Omega_{*,2} & \sim & - M_{*} \frac{GM_{\rm *}}{r_{\rm h}} \ = \ {\rm
    LSF}\ \Omega_{*,1}. 
\end{eqnarray}

Now we calculate the escape velocity of the system after the gas is
gone as
\begin{eqnarray}
  \label{eq:vesc1}
  v_{\rm esc} & \sim & \sqrt{-\frac{2}{M_{*}} \Omega_{*,2}}.
\end{eqnarray}
If we now replace $\Omega_{*,2}$ by LSF times $\Omega_{*,1} =
-T_{*,1}/Q_{\rm f}$ we have
\begin{eqnarray}
  \label{eq:vesc2}
   v_{\rm esc} & = & \sqrt{3\kappa} \sqrt{\frac{\rm LSF}{Q_{\rm f}}} \sigma_{*}.
\end{eqnarray}

A reasonably first guess of $f_{\rm bound}$ would be the fraction of stars with velocities 
below the escape velocity. If we assume a Maxwellian velocity distribution, then  $f_{\rm bound}$  
 is given by its cumulative probability distribution with the form:
\begin{eqnarray}
  \label{eq:maxwell}
  F( < X ) & = & {\rm erf}\left( \frac{1}{\sqrt{2}} X \right) -
  \sqrt{\frac{2}{\pi}} X \exp \left( - \frac{X^{2}}{2} \right)
\end{eqnarray}

where $X = v_{\rm esc}/a$. Since $a^2=\kappa\sigma_*^2$ then $X=v_{\rm esc}/\sqrt{\kappa}\sigma_*$ and finally:
\begin{eqnarray}
  \label{eq:fbound}
  f_{\rm bound} & = & {\rm erf} \left( \sqrt{\frac{3}{2} \frac{\rm LSF}{Q_{\rm f}}} \right ) - \sqrt{ \frac{6}{\pi} \frac{\rm LSF}{Q_{\rm f}} } {\rm exp} \left ( -\frac{3\rm LSF^2}{2Q_{\rm f}^2}
  \right ).
\end{eqnarray}

In Fig.~\ref{fig:extremes} we show $f_{\rm bound}$ against LSF for
various values of $Q_{\rm f}$.  The solid black line is the fit from
above which has no free parameters.  This simple model describes 
the data points of our simulations very
well, especially if we look at high LSF and
low $Q_{\rm f}$ values, i.e.\ when we do not lose many unbound stars
(upper panels).  

When we have high $Q_{\rm f}$ values as in the lower panels of
Fig.~\ref{fig:extremes} the simple model (solid black line) tends to
over-estimate the final bound fraction. We can apply a simple
correction. Following  the first estimate of $f_{\rm bound}$ a fraction
of stars is lost very rapidly after gas expulsion, and so the
escape velocity falls by a further factor $\sqrt{f_{\rm bound}}$ in
Eq.~\ref{eq:vesc2}.  We then have to solve
Eqs.~\ref{eq:vesc2} and~\ref{eq:maxwell} iteratively which gives the
blue dashed-lines in Fig.~\ref{fig:extremes}.  In most cases the true
values of $f_{\rm bound}$ are enclosed between the solid black and 
blue dashed-lines suggesting that reality is somewhere inbetween.

We have constructed a simple analytic approximation with no free
parameters that estimates the final bound fraction from the values of
the stellar virial ratio and LSF at the moment of gas expulsion.
Given the simplifying assumptions we have made it is very gratifying
that this seems to explain the results so well.

\subsection{The effect of the gas potential}
\label{sec:gaspot}

In Fig.~\ref{fig:extremes} points are coloured according to the form
of the gas potential:  blue is a Plummer potential, and red a uniform sphere.
There appears to be a very strong dependency on the form of the gas
potential.  In Fig. 3 systems with concentrated gas potentials
($R_{\rm pl} = 1$~pc) shown by the blue markers are concentrated to the
left of each panel with low LSF and low $f_{\rm bound}$.  Systems
with extended gas potentials ($R_{\rm pl} = \infty$) shown by the red
markers are towards the right with higher LSF and $f_{\rm bound}$.

Taken at face value this suggests that the form of the gas potential
is crucial in determining the fate of a system.  However, this is not
the case. Rather it is due to a link between the form of the gas
potential and the possible values of the LSF.  The LSF measures the
relative masses of gas and stars within the half-mass radius of the
stars.  Gas outside this radius is not taken into account.  Even
though the total mass in gas in the whole star forming region stays
constant, the LSF fluctuates as the half-mass radius of the stars
fluctuates (this is the motivation for the introduction of the LSF by
 \cite{Smith2011a}.

In a bound, fractal distribution the stars can do nothing except
collapse to a denser (and smoother) configuration.  Much of the
initial potential energy in a fractal distribution is localised in
substructure and is redistributed during violent relaxation.  The
potential energy, $\Omega$, of a system is 
\begin{eqnarray}
  \label{eq:potfrac}
  \Omega & \sim &- A \frac{GM^{2}}{R}
\end{eqnarray}
where $M$ is the mass of the system and $R$ some characteristic radius
(and $G$ the gravitational constant). $A$ is a measure of the mass
distribution of the system. For a Plummer sphere, if $R$ is the
Plummer radius then $A \sim 0.3$.  But for a $D = 1.6$~fractal, when
$R$ is the initial size of the system, $A \sim 1.5$.  Therefore, the
violent relaxation of a fractal causes a significant decrease in the
size of the system (see \citealt{Allison2009} for details).

\begin{figure}
    \begin{center}
      \includegraphics[width=3.2in]{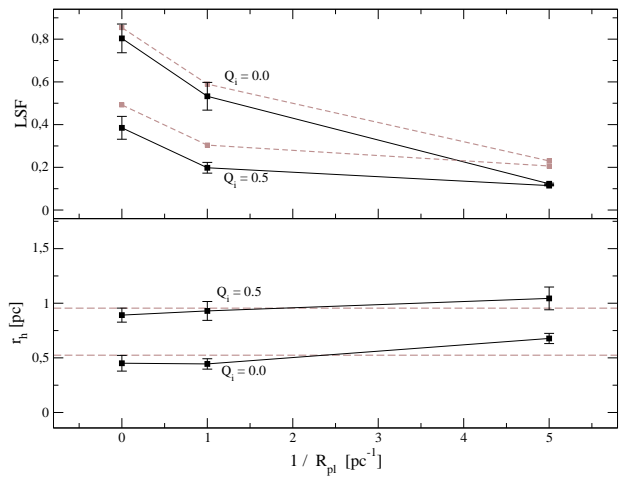}
    \end{center}
    \caption{The variation of the LSF of the clusters due to the change in the
      concentration of the background gas. Top panel: The average of the LSF of the simulations against the inverse of their scale lengths. A black solid line connects simulations with the same initial virial ratio as labelled. Bottom panel: The half mass radius is only weakly dependent on the gas scale length. The average is shown by the horizontal dashed line. The brown dashed line in the upper panel is the recalculated LSF using a fixed half-mass radius with the average value. }
    \label{fig:background}
  \end{figure}

Exactly how such a system will contract depends on the exact details
of the initial fractal distribution, the initial virial ratio ($Q_{\rm
i} = 0$ systems will contract more than $Q_{\rm i} = 0.5$ systems) and
how relaxed the system has become.  However, we find it does
not depend on the shape of the background gas potential, as shown in Fig.~\ref{fig:background}. In the upper panel, symbols with error bars are the average LSF of simulations at the moment of gas
expulsion. We include data points for (from left to right) $R_{\rm pl} = \infty$ (uniform gas), $R_{\rm pl}= 1$~pc, and also $R_{\rm pl} = 0.2$~pc (a very concentrated gas distribution). On the
x-axis, we plot $1/R_{\rm pl}$ in order to place all simulations on the same plot. There are two curves for the two different initial virial ratios ($Q_{\rm i} = 0.0$ and 0.5). There is a clear trend
for the LSF to be lower as the gas becomes increasingly concentrated. To understand this, we must bear in mind that the LSF is a function of the total gas mass within the half mass radius of the stars. Therefore a change in LSF could arise from either a change in the amount of gas surrounding the stars, and/or a change in the half-mass radius of the stars as the gas scale length is varied. We find that the half-mass radius is only a very weak function of the gas scalelength as shown in the lower panel of Fig.~\ref{fig:background}. Here symbols with error bars are the average half-mass radius $R_{\rm h}$ of the stars. This weak dependency demonstrates that the strong dependency of the LSF on gas scalelength arises mainly for the following reason -- by making the gas more concentrated, more gas is being placed about the stars, and the LSF is lowered. 

To confirm that the small variation in $R_{\rm h}$ with gas scalelength does not play a strong role, we calculate the average $R_{\rm h}$ for each set (see horizontal dashed lines in the bottom panel of Fig.~\ref{fig:background}). Now we fix $R_{\rm h}$ to have the average value (i.e a constant value for all gas scalelengths) and recalculate the LSF values at their new half-mass radius. The results are indicated by the brown dashed lines in the upper panel. The trend of LSF with gas scalelength is very similar, even when the stellar half-mass radius is fixed to be constant. This confirms that the strong dependency of the LSF on gas scalelength arises almost entirely for the following reason. Increasing the gas concentration places more gas about the stars, and does not change the stellar distribution significantly.

\section{Discussion and Conclusions}

Initially clumpy and irregular distributions of stars cannot be 
in dynamical equilibrium. As a result, they undergo violent relaxation
with initially significant changes in their virial ratio as they
expand and collapse, attempting to approach equilibrium. This 
occurs even when the clusters are initially `virialised'
(ie. $Q_{\rm i} = 0.5$).  These deviations are largest for very young
star clusters, and decrease as the cluster settles down, as
substructure is erased. As a result, the effects of gas expulsion at early times,
before the system has relaxed, depend strongly on the instantaneous
value of the virial ratio as well as the Local Star Fraction (LSF,
relative distribution of stars in the gas potential).

At later stages ($>$2 crossing-times), it is known that the LSF becomes the key predictor of cluster survival from gas expulsion, with second-order 
modifications due to the cluster's dynamical state (\citealp{Smith2013a}). However at these early stages when oscillations in the virial ratio are 
so large, we have shown that the dynamical state of the cluster may actually be equally influential (if not more influential) than the LSF.

A primary goal of studying the response of young, embedded star clusters to gas expulsion is to predict how well a cluster survives gas expulsion, based on its pre-gas expulsion properties. This study reveals that both the LSF, and the dynamical state can be important parameters dictating cluster survival to gas expulsion. Fortunately in our numerical studies, we can ascertain the exact value of the LSF and virial state. However, observationally, it may be incredibly challenging to measure either of these 
properties accurately. It is not inconceivable that the LSF might be calculated approximately by deprojection, although it would need to be a cluster 
caught very close to the instant of gas expulsion, or the LSF may later change. However, measuring the virial ratio of a real cluster is a huge challenge.

To worsen matters, our study reveals that in certain circumstances, even with a knowledge of both the LSF and virial ratio, the cluster survival maybe poorly constrained. For example, take a cluster with a low virial ratio (e.g $Q_{\rm f}$=0.34 at gas expulsion; upper-left panel of Fig.~\ref{fig:extremes}). If the cluster has an LSF$\sim$0.2 (a reasonable physical value), the $f_{\rm bound}$-LSF trend rises very steeply. Such a cluster is equally likely to be near destroyed (have $\sim90\%$ of its stars unbound),
as only weakly affected (losing $\sim30\%$ of its bound stars). Thus it is possible that, even if the virial ratio were measured, the result could place the cluster in a region of parameter space where the cluster survival 
could be anything from weak mass loss to near total destruction.

Comparing the panels of Fig.~\ref{fig:extremes}, we can see that clusters with LSF$\sim$0.2-0.4 are the most sensitive to their dynamical state. In comparison clusters with high LSF vary their resulting bound fractions very little, even for large changes in dynamical state. If LSF$\sim$0.2 is a typical value, then these results suggest that clusters which are observed post-gas expulsion, must have been sub-virial to avoid losing a large fraction of their stellar mass during the process.

Clearly our models are extremely simple conceptionally. They lack a large number of physical processes that are also highly important in young star clusters.
For example, our cluster stars have no initial mass function, we start our simulations with no binaries, we do not consider stellar evolution, and our treatment 
of the gas is highly simplified. Nevertheless, the use of such simple idealised models has enabled us to clearly determine the significant role of clumpy substructure and 
the dynamical state of the clusters on cluster survival following gas expulsion, through the use of controlled numerical experiments. This approach has revealed 
just how sensitive star clusters are to their dynamical state when gas expulsion occurs. We therefore suggest that real star clusters will be very sensitive,
perhaps as sensitive as our model star clusters, to their dynamical state when the gas is expelled at early times.

Our key results may be summarised in the following:
\begin{enumerate}
\item For early gas expulsion (before 2 crossing-times) we find the dynamical state of our model star clusters, measured at the time of gas expulsion, plays a key role in influencing cluster survival following gas expulsion. Star clusters may be highly super- or sub-virial in these early phases.
\item We show how the $f_{\rm bound}$-LSF trend can be well approximated with a very simple analytical model. The model matches the simulations best when the dynamical state is not extreme (i.e highly super- or sub-virial).
\item Clusters which have LSFs in the range 0.2-0.4 (physically reasonable values) are most sensitive to the virial ratio at the instant of gas expulsion. \item Clusters with low virial ratio have a very steep rise in the $f_{\rm bound}$-LSF trend. For such a cluster with an LSF$\sim0.2$, it is therefore not possible to predict if the cluster will be heavily destroyed or only mildly affected -- even knowing both the LSF and the virial ratio.
\end{enumerate}

This study highlights the difficulties faced in trying to determine the survival rate(s) of real star clusters due to gas expulsion. At early times, 
the dynamical state of a cluster may be far from dynamical equilibrium, and this can significantly affect the clusters survival to gas expulsion.
Thus a best estimate of a cluster's survival is found measuring both the LSF and virial ratio. Accurately measuring these two parameters for a real 
cluster represents a huge observational challenge, in particular the dynamical state. Furthermore, some clusters may be situated in regions of parameter space where their survival to gas expulsion remains highly uncertain, even knowing both the LSF and virial ratio.

\section*{Acknowledgements} 
JPF acknowledges funding through FONDECYT regular 1130521 and a 
CONICYT Magister scholarship. RS was supported by the Brain Korea 21 Plus Program (21A20131500002) 
and the Doyak Grant (2014003730). RS also acknowledges support from the EC through an ERC grant
StG-257720, and FONDECYT postdoctorado 3120135. MF is funded through 
FONDECYT regular 1130521. GNC acknowledges funding through 
FONDECYT postdoctorado 3130480. MB is funded through a CONICYT 
Magister scholarship, BASAL PFB-06/2007 CATA and DAAD PhD scholarship . 

\bibliography{bibfile}
\bsp
\label{lastpage}

\end{document}